# Hyperbolic Transverse Patterns in Nonlinear Optical Resonators

Kestutis Staliunas[(1)] and Mustapha Tlidi[(2)]

[(1)] *ICREA, Departament de Fisica i Enginyeria Nuclear, Universitat Politecnica de Catalunya, Colom 11, E-08222 Terrassa, Barcelona, Spain.*
[(2)]*Optique Nonlinéaire Théorique, Université Libre de Bruxelles, Campus Plaine, CP 231, B-1050 Bruxelles, Belgium.*

**Abstract**

We consider a nonlinear optical system in general, and a broad aperture laser in particular in a resonator where the diffraction coefficients are of opposite signs along two transverse directions. The system is described by the hyperbolic Maxwell-Bloch equations, where the spatial coupling is provided by the D'Alambert operator rather than by the Laplace operator. We show that this system supports hyperbolic transverse patterns residing on hyperbolas in far field domain, and consisting of stretched vortices in near field domain. These structures are described analytically by normal form analysis, and investigated numerically.

Spontaneous pattern formation is a universal behavior observed in a diverse class of far from equilibrium systems in chemistry, biology, fluid mechanics, and nonlinear optics [1,2]. Such patterns are characterized by an intrinsic scale determined by the dynamical parameters and not by the geometrical constraints imposed by the boundaries. The formation of spatial structures is attributed to the balance between nonlinearities (chemical reaction or light-matter interaction) and transport process (diffusion and/or diffraction). Spatial structures in nonlinear optics have been discussed in a number of reviews [3-5]. All known pattern forming systems are governed by partial differential equations of elliptic type, *i.e.*, the spatial differential operators are of the Laplace type $\hat{L} = \left(d_{xx}\partial^2/\partial x^2 + (d_{xy}+d_{yx})\partial^2/\partial x \partial y + d_{yy}\partial^2/\partial y^2\right)$, where $\hat{d} = \begin{bmatrix} d_{xx} & d_{xy} \\ d_{yx} & d_{yy} \end{bmatrix}$ is the diffraction matrix of real valued elements with the determinant $|\hat{d}| > 0$. Under an appropriate coordinate transformation the elliptical spatial differential operator can be rewritten into the form $\hat{L}_{1,1} = \left(\partial^2/\partial x^2 + \partial^2/\partial y^2\right)$. In optical systems, the light propagation can be affected by both diffusion and/or diffraction, *i.e.*, the coefficient at differential operator is real and/or imaginary, respectively. The diffusion is always governed by an elliptical type of operator, whereas the diffraction process can either be modeled by an elliptic or a hyperbolic type of operator. In the latter case $|\hat{d}| < 0$, and a transformation of coordinates leads to the D'Alambert type of operator $\hat{L}_{1,-1} = \left(\partial^2/\partial x^2 - \partial^2/\partial y^2\right)$. Physically, this means that diffraction coefficients in the resonator have opposite signs for different transverse directions. This can occur *e.g.,* in the photonic crystal-like resonators, the resonator with periodic modulation of refractive index along one transverse (say *y*) direction. The periodic spatial modulation of refraction index allows to modify, to reduce, or even to invert the sign of the diffraction coefficient. This phenomenon has been recently studied in conjunction with so called band-gap solitons in nonlinear optics [6], in atomic Bose-Einstein condensates [7], and in dissipative nonlinear photonic crystal resonators [8]. More recently, an experimental evidence of the inversion of diffraction coefficient sign in one-dimensional optical system has been reported in [9]. We note the recent work [10] where the periodic modulation of potential in one spatial direction resulted in a reversal of the sign of the diffraction coefficient (effective mass) in that direction, whereas the diffraction coefficients in other directions remained unchanged.

Another example of hyperbolic systems is the near self-imaging resonators that also allow manipulating the value and the sign of the diffraction coefficient. In particular, the diffraction coefficient can be tuned exactly to zero at precisely self-imaging configuration [11]. Such resonators are convenient for experimental observations of nonlinear optical patterns [12]. For cylindrical (astigmatic) resonators the net diffraction coefficient can be manipulated separately for different transverse directions. Consequently the differential operator $\hat{L}_{1,1}$, which is an elliptical Laplace operator for spherically symmetric resonators, can be converted into hyperbolical D'Alambert operator $\hat{L}_{1,-1}$ for cylindrical resonators.

Despite of the importance of the hyperbolic patterns that result from the competition between hyperbolic diffraction and nonlinearity, they have neither been experimentally observed nor theoretically investigated. The purpose of this Letter is to fill this gap by investigating the theoretical side of this problem. We illustrate the importance of this competition by studying the formation of transverse structures in the full Maxwell-Bloch equations (MBE) under the effect of hyperbolic diffraction. In particular, our study reveals that the hyperbolic MBE support transverse patterns residing on hyperbolas in far field domain, and respectively consisting of stretched vortices in near field domain. These predictions are confirmed by an analytical description based on the normal form reduction. The normal form approach makes our analysis more general: one can expect that hyperbolic patterns constitute a dominant dynamical behavior in a general class of nonlinear optical systems, and not only in a broad aperture laser systems.

We consider the single longitudinal mode large aperture lasers. In the good cavity limit, the spatiotemporal evolution of the slowly varying envelopes of the electric field E, polarization P, and population inversion D, can be described by the hyperbolic MBE:

$$\frac{\partial E}{\partial t} = -(1+i\omega)E + P + i\hat{L}_{1,-1}E \tag{1.a}$$

$$\frac{\partial P}{\partial t} = -\gamma_\perp (P - ED) \tag{1.b}$$

$$\frac{\partial D}{\partial t} = -\gamma_{II}\left(D - D_0 + \frac{EP^* + E^*P}{2}\right) \tag{1.c}$$

Where $\hat{L}_{1,-1} = \partial^2/\partial x^2 - \partial^2/\partial y^2$ is the D'Alambert operator acting in the transverse plane (x, y). Time *t* is scaled to the photon life time $\kappa$, $\omega$ is the normalized detuning parameter (mismatch between atomic gain and cavity resonant frequency normalized to the cavity line-width), $\gamma_\perp$ and $\gamma_{II}$ are the decay rates of the polarization and the population inversion normalized to photon decay rate respectively. $D_0$ is the saturated population inversion. The diffraction coefficients in (1.a) are normalized to unity by proper scaling of the spatial coordinates *x* and *y*.

The equation (1) differs from the usual (elliptical) MBE system [13] in the spatial differential operator only. The MBE (Eq. 1) applies in various physical situations, *e.g.*, periodic modulation of the refractive index in the y direction or cylindrical astigmatism of the resonator. In the first of two situations mentioned above, when the origin of the hyperbolic type of diffraction is due to periodic modulation of the refraction index in one direction, the derivation is performed along the lines described in [10]. Starting from the usual MBE [13] plus a periodic modulation of refractive index in one (y) transverse direction, one expands the electromagnetic field into a set of spatial modes (i.e., spatial harmonics) and keeps only the two most relevant ones. One then computes the corresponding eigenvalues and eigenfunctions (Bloch modes). Finally, rewriting the full nonlinear system in the basis of these Bloch modes, a differential operator of the D'Alambert type naturally appears for the envelope of the Bloch mode at the band-edge. In this derivation, one assumes that the amplitude of the harmonic modulation of the refraction index is small, which justifies the truncation of the expansion to only two spatial harmonics.

In the second situation, when the origin of the hyperbolic type of diffraction is due to cylindrical astigmatism of the near self-imaging resonator, the derivation is based on the ABCD matrix of the (astigmatic) resonator, and leads directly to the D'Alambert type of differential operator in (1.a). Semiclassical calculation of the coupling between the electric field with the matter, as performed in a standard way (see e.g. [13]) results in the hyperbolic MBE (1). A detailed derivation of the hyperbolic Maxwell-Bloch equation in both cases will be reported elsewhere.

The homogeneous steady states solutions of (1) are identical to those of the usual MBE. An essential difference appears, however, when considering pattern formation is that the spatial coupling between different points in the transverse plane is provided by the D'Alambert operator rather than by the Laplace operator. This spatial coupling will modify, as we shall see, the process of pattern selection that takes place in a resonator. In order to study transverse patterns in hyperbolic system, we performed a derivation of the order parameter equation, namely the hyperbolic complex Swift-Hohenberg equation (CSHE), and the modified complex Newell-Whitehead-Segel Equation (NWSE). These model equations are paradigmatic ones for the description of pattern formation process. The validity domain of the hyperbolic CSHE is limited by the following assumptions: i) "near threshold" condition: $p = (D_0 - 1) = O(e^2)$, where $p$ is distance from the first laser threshold; ii) "close to resonance" condition: $\Delta w = (\nabla^2 - w) = O(e)$; iii) class-C laser assumption: $g_\perp, g_{II} = O(1)$, where $e$ is a small parameter, which measures the distance from the first laser threshold. In addition, we consider slow space $(x, y) \equiv e^{1/2}(x, y)$ scales and two slow time scales related with the diffraction of the fields in resonator $t_1 = e\,t$, and with the build up time of the field in resonator: $t_2 = e^2 t$. This procedure has been used to derive the CSHE for the usual elliptical class-A lasers in [14] by adiabatic elimination of the slow atomic variables ($P$, and $D$), or in [15] by multiple time-space scale expansion, and for a laser with injected signal [16]. A multiple space-time expansion leads the following CSHE that included the hyperbolic nature of diffraction

$$\frac{\partial \Psi}{\partial t} = \left( i(\hat{L}_{1,-1} - w) - \frac{1}{(1+g_\perp)^2}(\hat{L}_{1,-1} - w)^2 + p - |\Psi|^2 \right) \Psi, \tag{2}$$

The order parameter $\Psi(x,y,t)$ is proportional to the amplitude of electric field, and the normalized time is $t = t g_\perp / (1 + g_\perp)$.

The linear stability analysis of the trivial zero solution of (2) (and obviously of the full MBE) gives some insight into the initial (linear) stage of the pattern formation in the system. The growth exponents of the tilted waves $\Psi(x,y,t) = \exp[\lambda t + i k_x x + i k_y y]$ are obtained straightforwardly from (2), and read:

$$\lambda(k_x, k_y) = p + i\left(-k_x^2 + k_y^2 - w\right) - \frac{1}{(1+g_\perp)^2}\left(-k_x^2 + k_y^2 - w\right)^2 \tag{3}$$

The maximum growth rate in Eq. (3) is given by condition $\partial \lambda_{Re}/\partial k_x = \partial \lambda_{Re}/\partial k_y = 0$, and occurs when $-k_x^2 + k_y^2 - w = 0$, which is an hyperbola in the Fourier space, where $\lambda_{Re}$ is the real part of the growth exponent $\lambda = \lambda_{Re} + i\lambda_{im}$. The trivial solution is unstable with respect to fluctuations with a wave-vector $\mathbf{k} = (k_x, k_y)$, thus there exists a finite band of linearly unstable Fourier modes given by $\left(-k_x^2 + k_y^2 - w\right)^2 < p(1+g_\perp)^2$. These unstable modes trigger the spontaneous evolution of the field, polarization and population difference towards spatial patterns which occupy the whole space available in the transverse plane of the cavity. This simple relation illustrates a remarkable property of the linearly unstable area, which contrary to the usual elliptic CSHE where unstable modes form a ring in the Fourier space, is here bounded by hyperbolas. This behavior is displayed in the Fig.1. The hyperbolic shape of the instable area has others important implications in the pattern forming process: i) in contrast with the elliptic case with no preferred direction because of the isotropy in the $(x,y)$ space, the rotational symmetry is broken in the systems with hyperbolic diffraction, and therefore the emerging pattern depends strongly on the direction of underlying tilted wave $\mathbf{k} = (k_x, k_y)$; ii) wave-vectors with different magnitudes can be simultaneously excited. This means that there is no well defined critical wave-number as in elliptical systems. This absence of single critical wave-number is illustrated in the Fig.1 where, for instance, the possibility of patterns with two different wave-numbers $|\mathbf{k}_1|$ and $|\mathbf{k}_2|$ is shown.

Next we derive the amplitude equation, by applying the method of the separation of variables, i.e., we assume that the hyperbolic CSHE Eq. (2) has a solution of the form: $\Psi(x,y,t) = f(x,y,t)\exp(ik_x x + ik_y y)$. Substituting this solution into the Eq. (2), we obtain the amplitude equation:

$$\frac{\partial f}{\partial t} = i\hat{L}_{1,-1} f - 2\bar{\mathbf{k}}\nabla f + \frac{1}{(1+g_\perp)^2}\left(i\hat{L}_{1,-1} - 2\bar{\mathbf{k}}\nabla\right)^2 f + (p - |f|^2)f \qquad (4)$$

where $\bar{\mathbf{k}} = (k_x, -k_y)$ is the vector adjoined to wave-vector $\mathbf{k} = (k_x, k_y)$, in the sense that the $y$ components of both vectors are of opposite signs, and $\nabla$ is the gradient operator acting in $(x,y)$ transverse plane.

Equation (4) is similar to the Newell-Whitehead-Segel Equation (NWSE), universally describing slow spatial modulation of the stripe patterns in usual (elliptical) pattern forming systems [17]. The relevance of the complex NWSE in nonlinear optics has been established [18]. However, Eq. (4) differs from the usual real or complex NWSE derived for elliptical systems in several aspects: i) the obvious difference with elliptical NWSE is the differential operator of hyperbolic type $\hat{L}_{1,-1} = \partial^2/\partial x^2 - \partial^2/\partial y^2$; ii) another peculiarity is that the advection in the system described by Eq. (4) occurs along the direction $\bar{\mathbf{k}} = (k_x, -k_y)$ (see the second r.h.s. term of Eq. (4)), and not along the direction of the underlying tilted wave $\mathbf{k} = (k_x, k_y)$ as could be intuitively expected; iii) the character of the pattern formation depends on the direction of underlying tilted wave. This property contrast strongly with the elliptical case due to rotational invariance of the elliptic CSHE. In hyperbolic case the rotational invariance is broken, as seen from both hyperbolic equations (2,3), and as illustrated in the Fig.1.

Next, we numerically check the predictions of the above analysis of order parameter equation (2) and amplitude equation (4) by solving numerically the full hyperbolic MBE (1). In order to ensure the stability of the solutions, we added phenomenologically a small diffusion in MBE, *i.e.*, the differential operator was used of the form $i\hat{L}_{1,-1} + g\hat{L}_{1,1}$ with $g \ll 1$. Physically the diffusion in optical systems corresponds to the spatial frequency filtering present in most systems, and can be considered small compared with the diffraction. In our numerical study, the diffusion/diffraction ratio is fixed to $g = 10^{-3}$, but the results depend only negligibly on this ratio. We consider a system of large size compared to the scale of the selected pattern. The characteristic size of the dissipative nonlinear optical system is characterized by the dissipative Fresnel number F, which for the MBE (1) with normalized parameters is equal to the square of integration region, and for our calculations is $F \propto 1000$. In terms of experimentally accessible parameters, the dissipative Fresnel number $F = 4\pi a^2/(L\lambda Q)$, ($a$ is the transverse size of the system, L – full resonator length, $\lambda$ - the wavelength and $Q$ is finesse of the resonator), which is equal to the usual Fresnel number divided by the resonator finesse. The dissipative Fresnel number is a measure of a maximal number of dissipative structures (*e.g.*, vortices) that can be placed independently (or weakly dependently) one from another within the aperture (or within the integration range with periodic boundaries). The dissipative Fresnel number was sufficiently large in our numerical calculations (F=900) thus the influence of used periodic boundaries was negligible. In this case, the distance between the closest eigenvalues of the linear problem is negligibly small, leading to the quasicontinuous spectrum the D'Alambert operator. We mention that in the opposite limit of for conservative system with $Q \to \infty$ (resulting in $F \to 0$) the boundaries influence critically the spatiotemporal light dynamics as shown in [19].

The profile of the field intensity in 2D Fourier space during the linear stage of evolution is shown in Fig.2 for different values of detuning. The plots are given at a normalize time of 20, which means that the field intensity is still not saturated (population inversion still not depleted). The far field patterns show formation of the hyperbola (Fig.2.a), as predicted. We note that the change the sign of detuning results in spatial orientation of the hyperbola as shown in the Fig.2.b. This is an essential difference with elliptical systems where transverse patterns occur for only one sign of detuning: since the maximum growth rate realizes for $k_x^2 + k_y^2 = -\omega$ for elliptical systems, no pattern formation is possible for positive detuning. For zero detuning hyperbola degenerates into two bisectrices $k_x^2 - k_y^2 = 0$, as shown in the Fig.2.c. The patterns can occur on all spatial scales in this singular case. For $\omega \neq 0$ patterns can occur on scales $|\mathbf{k}|^2 > |\omega|$.

The transverse profiles of hyperbolic pattern during the *nonlinear* stage of evolution are shown in Fig.3. These structures consist of a remarkably robust parabolic patterns composed by several stretched vortices advected by the underlying tilted wave. The underlying tilted wave in Fig.3.a is directed along the focus line of the hyperbola (corresponds to $\mathbf{k}_1$ in Fig.1), and in Fig.3.b is directed nearly along asymptotes of the hyperbola (corresponds to $\mathbf{k}_2$ in Fig.1). An essential difference with elliptical systems is that the vortices are typically strongly squeezed elliptic, and that the differently charged vortices have different intensity profiles.

The asymptotical cases shown in Fig.3.a and Fig.3.b can be treated analytically by amplitude equation. For case of Fig.3.a, when the background tilted wave is nearly along y-direction (corresponding to $\mathbf{k}_1$ in Fig.1), the NWSE (4) simplifies to:

$$\frac{\partial f}{\partial t} = i\left(\frac{\partial^2}{\partial x^2} - 2ik_y \frac{\partial}{\partial y}\right)f - \frac{1}{(1+g_\perp)^2}\left(\frac{\partial^2}{\partial x^2} - 2ik_y \frac{\partial}{\partial y}\right)^2 f + (p - |f|^2)f \qquad (5)$$

which is very similar to the complex NWSE as derived in [13], except for that the sign of the term $2ik_y \partial/\partial y$ (which is opposite here). This coincidence is indeed obvious, because the instability "banana" for hyperbola and ellipse are similar, but just differently oriented in spatial Fourier domain. The advection in the Eq. (5) follows in opposite direction, *i.e.*, antiparallel to $k_y$.

For case of Fig.3.b, when the background tilted wave is nearly along asymptotes of the hyperbola (corresponding to $\mathbf{k}_2$ in Fig.1), the NWSE (4) simplifies to:

$$\frac{\partial f}{\partial t} = -2\bar{\mathbf{k}}\nabla f + \frac{(2\bar{\mathbf{k}}\nabla)^2 f}{(1+g_\perp)^2} + (p - |f|^2)f \tag{6}$$

where the instability area extends along the asymptote, and the advection occurs in the direction perpendicular to the underlying tilted wave.

In summary, we predict hyperbolic light patterns in nonlinear resonators, and in particular in broad aperture lasers, resulting from the competition between hyperbolic diffraction and nonlinearities. They consist of patterns residing on hyperbolas in far field configuration, and consisting of stretched vortices in near field configuration. The normal form reduction of the problem makes our analysis general: one can expect that the hyperbolic pattern constitute a class of patterns in different nonlinear optical resonators such as optical parametric oscillators, photorefractive oscillators and others. A better understanding of hyperbolic patterns is therefore crucial to the advancement of knowledge of complex nonlinear systems.

The work has been partially supported by Sonderforschungsbereich 407 of Deutsche Forschungsgemeinschaft, by project FIS2004-02587 of the Spanish Ministry of Science and Technology, and by the Fonds National de la Recherche Scientifique (Belgium) and the Interuniversity Attraction Pole of the Belgian government.

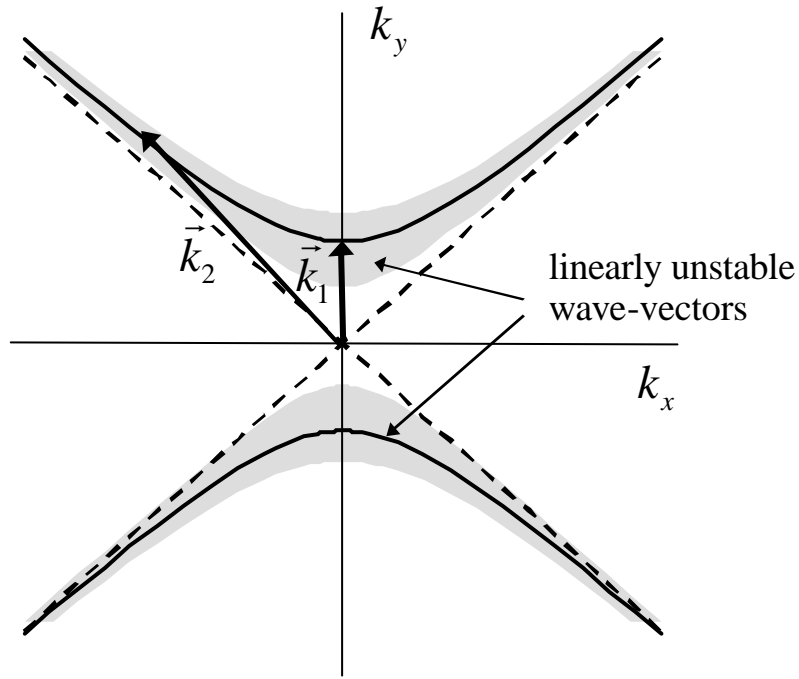

Fig.1. The maximum instability lines (solid), and the instability (grey) areas in spatial wave-vector space (far field domain) for hyperbolic pattern forming systems, as given by (2). The width of the instability area depends on gain parameter $p$. The plot corresponds to positive detuning $w$. For zero detuning the hyperbola degenerates into a cross-like area around the bisectrices $k_x \pm k_y = 0$; For a negative detuning the instability hyperbola is rotated by 90 degrees.

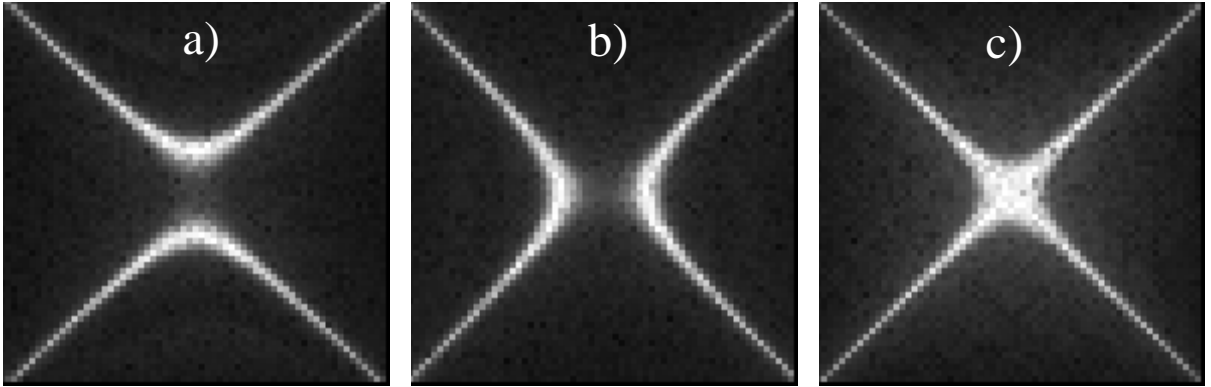

Fig.2. Distribution of field intensity in spatial Fourier domain (intensity of the far field) as obtained by numerical integration of hyperbolic MBE Eq.(1), for different detuning values. a,b) for $w = \pm 2$ resulting in far-field distributed around vertically (horizontally) oriented hyperbola; c) for $w = 0$ resulting in far-field distributed around asymptotes. The analytical maximum growth lines, as given by (2), are indicated by dashed lines. Other parameters are $D_0 = 2$, $g_\perp = 0.2$, and $g_\parallel = 1$. The size of integration region is 30. The integration time is 20, corresponding to a linear stage (population inversion still not depleted). The grid size is 128x128.

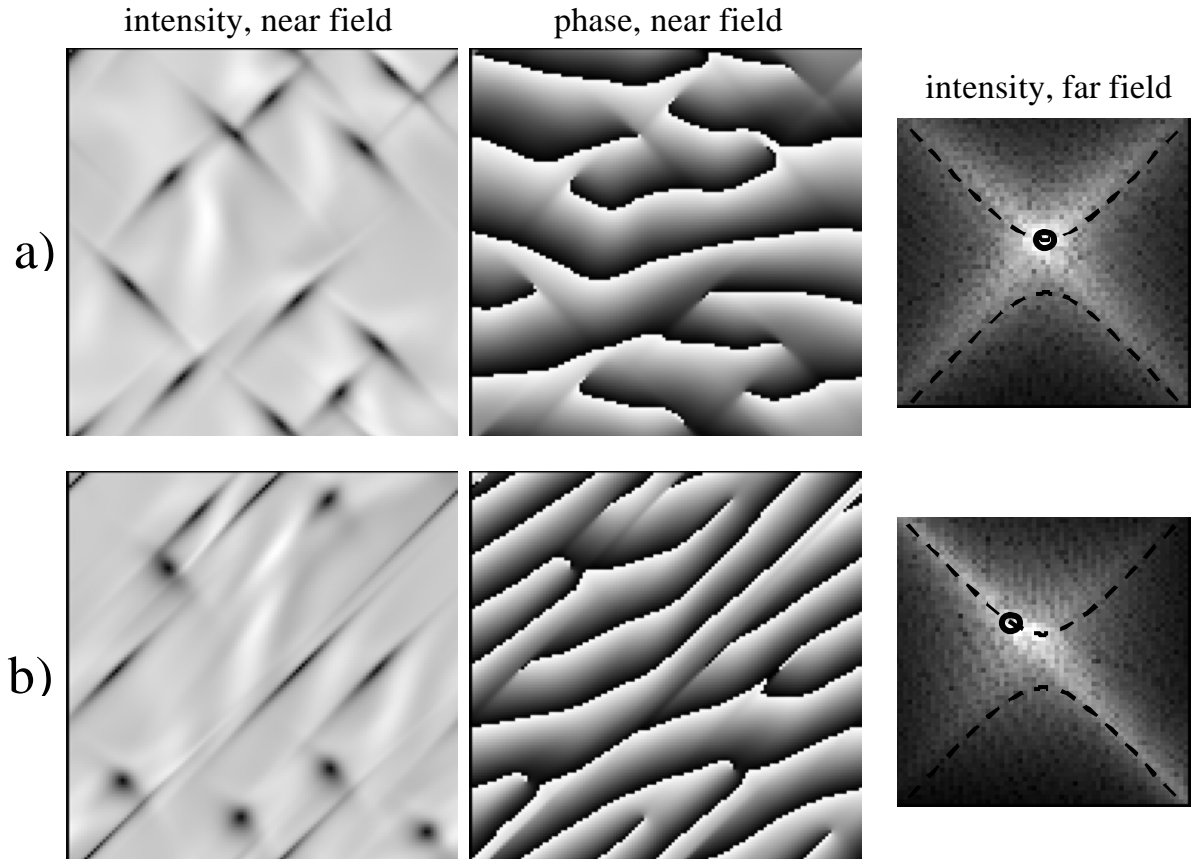

**Fig.3.** The intensity distribution (left), the phase distribution (middle), and the distribution in spatial Fourier domain (right), as obtained by numerical integration of hyperbolic MBE (1). The parameters are $D_0 = 2$, $g_\perp = 1$, $g_\parallel = 1$, and $w = 0.5$. The size of integration region is 50. The integration time is 200, which means that the essentially nonlinear stage is reached. The grid size is 128x128. The dashed lines in spatial Fourier domain correspond to analytically calculated maximum growth line; the circles indicate the tips of the background tilted waves. (a) Vortices for small positive detuning and when the background tilted wave is along y-direction (corresponding to $\mathbf{k}_1$ in Fig.1). Note that differently charged vortices are stretched in different directions, and stretching axes are at angle of almost 90 degrees. (b) Vortices for the case when the background tilted wave is nearly along the asymptote (corresponding to $\mathbf{k}_2$ in Fig.1). Note that differently charged vortices have different intensity distributions.